# A GEOMETRICAL APPROACH TO EVALUATING THE HEAT FLUX PEAKING FACTOR ON FIRST WALL COMPONENTS

R. Mitteau[1*], and P. Stangeby[2]

[1] Association Euratom - CEA sur la Fusion Contrôlée, direction des sciences de la matière, Centre d'Etude de Cadarache F-13108 Saint Paul Lez Durances CEDEX

[2] General Atomics , San Diego, California 92186-5608, US, and University of Toronto Institute for Aerospace Studies, 4925 Dufferin St., Toronto, M3H 5T6, Canada.

## Abstract

In magnetic fusion experiments, a simple technique to evaluate the heat flux on first wall components is a key to controlled plasma surface interaction. The heat flux can be characterized by the peaking factor which is the ratio of the peak heat flux to the average heat flux. The peaking factor can be calculated exactly using simple derivations and standard software tools. This analysis is applied to an Iter class experiment for plasma wall contact during start up phases at 15 MW, in idealised, realistic and misaligned situations. Even though the peaking factors are usually above 10, the peak heat load on the wall remains moderate at a few MW/m².

* corresponding author

# 1.  Introduction

Estimating the peak heat load ($q_{max}$) on plasma facing components (PFCs) before plasma operation is a key safety issue of plasma surface interaction analysis. Various methods are possible and can be categorised according to their principles. More scientific techniques start from the plasma characteristics (temperature, density, transport, sheath transmission coefficient) and then express the heat flux to the component. These methods are often limited by the imprecision of the plasma characteristics and especially the transport coefficients. The most elaborated derivations make use of edge plasma codes, assuring a high level of self consistence [1-3]. Accounting for the detailed wall connections eventually requires a 3D mapping of the SOL [4], allowing a rather comprehensive description of small scale plasma variations.  A completely different technique uses the monte carlo principle [5] but is of little practical use as it requires extensive calculation time for actual 3D wall surfaces. Other authors have used semi-analytic derivations [6,7] which are dependant on the experiment and only allow an approximate accounting of the cross component shadow. The most elementary method consists in dividing the exhaust power ($P_{total}$) by the component area (A), which gives an average heat load ($q_{mean} = P_{total}/A$), and then applying a peaking factor ($P_f$), $q_{max} = P_f \times q_{mean}$. This simple technique based on the peaking factor is valuable because few parameters are needed. However most of the useful information is hidden in $P_f$, which is too often just estimated or guessed. The peaking factor $P_f$, can however be calculated exactly, using simple scrape of layer (SOL) assumptions (exponential heat flux decay in the SOL : $\lambda_q$, cosine law and shadowing). Using field line calculations, the shadow can be precisely calculated. The influence of the plasma properties enters through, $\lambda_q$, which is often documented (scaling laws [8]). This technique will be explained in section 2 and applied to an Iter class experiment for illustration.

## 2.  Peaking factor calculation.

The technique requires a mesh of the surface, described by a set of elements $\{E\}_{\mathcal{C}}$ based on a set of nodes $\{N\}_{\mathcal{C}}$ (Fig. 1), defined for the investigated component $\mathcal{C}$. The technique also requires data for the magnetic configuration, given as the three components of the magnetic field (Br, Bz, Bφ). On each node n of $\{N\}_{\mathcal{C}}$, the relative heat flux magnitude j(n) is evaluated using the cosine law [9,10], possibly adding a cross field heat flux fraction [11]. As a result, the heat flux profile on the component is known, although in arbitrary units (fig. 2.a). The actual heat flux entering the component on each node n is $q_{sep}^{//} \cdot j(n)$, where $q_{sep}^{//}$ is the parallel heat flux density on the last closed surface. This picture is rendered more complex by the possible shadowing between components, caused by SOL connections. They are evaluated for each node n for $\mathcal{C}$ by field line tracing (fig. 1) to another component (or more) representing the rest of the wall (the shadowing component $\mathcal{S}$), represented by a mesh ($\{E\}_{\mathcal{S}}$, $\{N\}_{\mathcal{S}}$), fig. 1. In the simplified scheme used here, the heat flux is set to zero if a connection is found (full shadowing, c(n) = 0) and c(n) is set to 1 if the field line escapes $\mathcal{S}$ (fig. 2.b). The integral of c(n) over the component gives the wetted area.

The numerical projection of j(n) onto the elements e of $\{E\}_{\mathcal{C}}$ allows calculation of a surface integral, giving a relation between P$_{total}$ and $q_{sep}^{//}$ : $P_{total} = \iint_C q_{sep}^{//} \cdot j(p) \cdot c(p) \cdot ds$.

$q_{sep}^{//}$ is assumed constant for each node n of the component surface and can be factorised. With discrete formulation, the integral is expressed as a sum and $q_{sep}^{//}$ can then be obtained from :

$$q_{sep}^{//} = \frac{P_{total}}{\sum_{e \in \{E_C\}} j(e) \cdot c(e) \cdot ds(e)}$$

The denominator is obtained using a standard software operator. It has the dimension of a surface and is thus an effective area. As it relates the power to the separatrix heat flux, it can be referred as "parallel heat flux density effective area" ($A_{sep}^{eff}$). Having quantified $q_{sep}^{//}$ gives access to the absolute value of the heat flux everywhere at the surface of the component. The peak heat flux on the component $q_{max}$ is the maximum value of the product $q_{sep}^{//} \cdot j(n) \cdot c(n)$ over all nodes $n \in \{N\}_e$. The peaking factor Pf is then obtained through Pf = $q_{max} \times$ A / $P_{total}$.

This technique is applicable to a set of limiters which are perfectly aligned. In this case, the calculation is simplified to a single element of the component, assuming toroidal symmetry. If N is the limiter number, the method is applied to a single limiter element using a power of $P_{total}$/N. The method is also applicable to misaligned limiters, the only price to pay being the loss of toroidal symmetry, with the consequence of much larger meshes and longer run times.

## 3. Results for various limiter situations

The model is applied to representative wall situations in an Iter size experiment. A 6.5 MA limiter plasma is used, the one at the end of the limiter phase in Iter (scenario 2 at 24.17s). For the modelling presented, the plasma continuously looses $P_{total}$ = 15 MW to the wall. The outer poloidal curvature radius of this plasma is a = 3 m and the magnetic pitch 7.9°.

The first results are for a series of sets of N outboard poloidal limiters, with N=9, 18 and 36 (a limiter is represented in fig. 1 with two neighbours at +/- 20°. The neighbours are reduced to their most forward ridge for simplification, an operation that has no influence of the shadow frontiers. The geometry of an individual PL is as follows : the intersection of one limiter with the equatorial plane is an arc, making a 2 centimetre bulge over a toroidal span of 1.5 m. This corresponds to a radius of 14 m in a horizontal plane. The profile is rotated along a horizontal axis in the poloidal direction, with a radius of 4.5 m matching approximately the outer poloidal curvature of the Iter wall. This results in a surface with a double curvature. The area is A = 10.6 m². The heat flux decay length is prescribed to be 10 mm, an average value representative of limiter phases in Iter. The number of poloidal limiters (PL) being high, the sink action of the PL set is that of a toroidal limiter, and $\lambda_q$ is kept constant. The heat flux pattern and the wetted area are calculated and the results given in table 1. The peaking factor ranges from 23 to 72, indicating that the peak heat flux is an order of magnitude above the mean heat flux. The high peaking factor is caused by a combination of factors : the poloidal curvature mismatch between limiter and plasma, small heat flux decay length and cross limiter shadowing. The wetted area is a significant fraction of the total limiter area (24% to 53%) : the highest set of limiters (36) has also the smallest wetted (i.e. unshadowed) area fraction (24%) because of the cross connections between individual limiter segments. Interestingly, the heat flux on the last closed flux surface is almost constant with the number of limiters N. Finally, the peak heat load has only as small dependence on the number of limiters. It can also be noticed that the change of $q_{sep}^{//}$ is only of 17% while the area changes of 400%. This is a confirmation that the sink action of the limiter set toward the plasma is well approximated by the one of a toroidal limiter. Should this not be the case, $q_{sep}^{//}$ would have a much bigger correlation to the limiter number.

In order to investigate the effect of the local geometry, a series is computed for five wall surfaces labelled 1 to 5 on the outside, whose results are given in table 2. These wall surfaces are made to occupy half of the outboard wall to account for 18 equatorial ports (one 10° sector is occupied every 20°). Case (1) is for a smooth wall made of an outboard toroidal surface. This case generates no shadow, and the heat pattern is formed of two toroidal bands. The peak heat flux is multiplied by a factor of 2 to account for the fact that only half of the wall is used. This is a limiting case, not possible technically, as shaping is required to care for the leading edges. It is however an lower ideal limit, toward which shape optimisation should tend. The heat flux is low (0.6 MW/m²) with a Pf of 7, the lowest of all case. Case (2) is a facetted wall : this wall consists of flat panels in front of each blanket module. The peaking factor jumps to 38 as the wetted area is reduced by shadowing. It is a significant difference to the ideal case, considering the limited surface change. Cases (3) and (4) are poloidal limiters.

(3) is circular in the poloidal direction, as in the previous section. The limiter labelled (4) is very close to (3), but is constituted of 1 m flat segments in the poloidal direction to match the shaping of the first wall modules. This is a shape that matches the current Iter shield modules, and is a significant design simplification compared to design (3). These two shapes give very similar results, with Pf ~ 37. This indicates that the price to pay for having straight panels in the poloidal direction is very small. Case (5) uses a spherical meniscus with a uniform curvature radius of 14 m, in both poloidal and toroidal directions. This case corresponds to first wall panel that are shaped in both toroidal and poloidal directions. These produce the highest peaking factor of 40. As in the previous section, the heat flux on the last closed flux surface is almost constant (+/- 20%) with the detailed surface shape. This is a remarkable result, considering the fact that the local geometry is changed of 5 cm, five time $\lambda_p$ : here also, this is a confirmation that the toroidal limiter approximation holds for such geometries.

The same technique can be used to investigate the effect of a misaligned component. This is tested on a inner cylindrical wall of radius R = 5m, of which a sector of 20° is supposed misaligned by a radial displacement of δ (Fig. 3). The curvature radius of the bulging module is adapted so that there is no leading edge. For the case without misalignment (δ = 0), the wall is actually a cylinder, and in that case, the peak heat load can be calculated analytically :

$$q_{max} = \frac{P_{total} e^{-1/2}}{2\sqrt{a\lambda_p}\, 2\pi R}$$

For a = 2.5 m, λp = 17 mm this peak heat load is 0.7 MW/m². The peaking factor is calculated to 3 by assuming a 2 m high wall, making a total surface of 63 m² (the mean heat flux is 15 MW / 63 m² = 0.24 MW/m²). With increasing δ, the peak heat flux and the peaking factor are evaluated using the numerical tool described in section 2. Both peak heat flux and peaking factor increase with δ, an intuitive result (table 3). On the regular surface, the peak heat flux decreases only slowly, from 0.7 to 0.45 MW/m². It is worth noting that a 2 mm misalignment (12% of λp) causes a two-fold increase of Pf, while a 20 mm misalignment (slightly above λp) creates a peaking factor increase of an order of magnitude. These results are influenced by the shape change with increasing misalignment (the incidence angle on the bulging module is variable), nevertheless they illustrate the effect of a misaligned element on the power distribution. This simple case can be used as a benchmark for the various modelling methods.

## 4.    Conclusion

Given the component geometry and the magnetic field configuration, the exact value of the peaking factor can be calculated numerically rather than being estimated or guessed. High values are found for poloidal limiters in Iter sized experiments (usually greater than 10). These values are caused by the mismatch of curvature radius, as well, of course, the peaking effect caused by short $\lambda_q$. $q_{sep}^{//}$ does not vary with the geometrical details of the wall, but rather with the large scale arrangement. This is a confirmation that for a sufficiently large number of discrete limiter-objects the plasma-wall contact is well approximated by a toroidal limiter type geometry, which simplifies the analysis of the wall. Local shaping will be required to take care of gaps, misalignments (as for the JET Iter-like wall [12]), and of the long toroidal span caused by the NBI ports in Iter [13]. Nevertheless, such an analyse show that input power, heat flux decay length and plasma scenario is a sufficient set of data to perform a detailed analysis of first wall geometries in tokamaks.

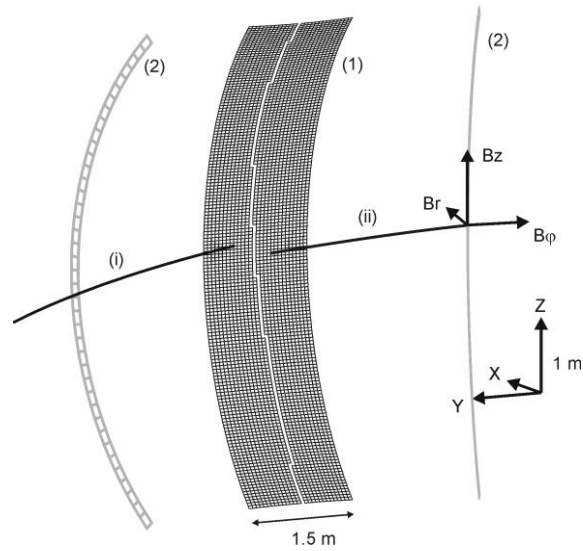

Fig. 1 : Component mesh Ecomp (1, in black) and shadowing mesh Eshad (2, in gray). Eshad is the central ridge of Ecomp, rotated from +/- 20° degrees. Two field lines are represented, (i) escapes the shadowing mesh, and the node of Ecomp from which it originates is designed as wetted, whereas (ii) connects to neighbour limiter, so that the the node of Ecomp from which it originates is designed as shadowed.

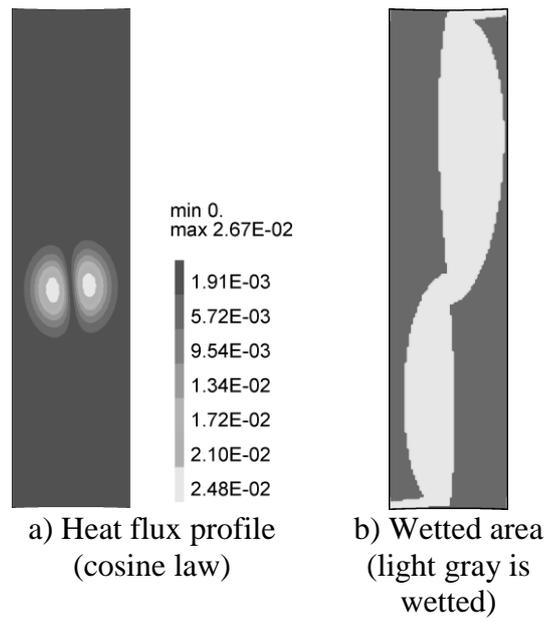

a) Heat flux profile
(cosine law)

b) Wetted area
(light gray is wetted)

Fig. 2 : Heat flux profile and wetted area for the limiter described in Fig. 1

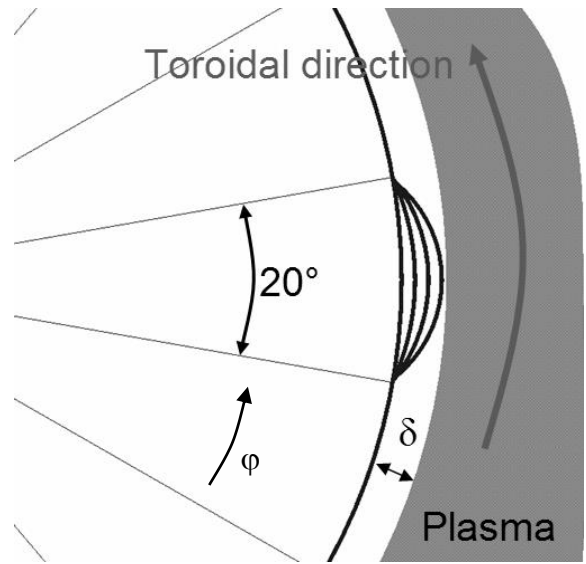

Fig. 3 : Wall and plasma cross sections in the equatorial plane with a misaligned module. For

$\delta = 0$, the wall is a pure cylinder.

| Number of limiters | 9 | 18 | 36 |
|---|---|---|---|
| toroidal span between limiters (°) | 40 | 20 | 10 |
| Total area (m²) | 96 | 191 | 382 |
| Wetted area (m²) | 51 (9 x 5.66) | 74 (18 x 4.12) | 100 (36 x 2.78) |
| $q_{sep}^{//}$ (MW/m²) | 133 | 116 | 113 |
| $q_{max}$ (MW/m²) | 3.6 | 3.1 | 2.8* |
| Peaking factor | 23 | 37 | 72 |

Table 1 : Results for 9, 18 and 36 outboard poloidal limiters for $P_{total}$ = 15 MW. In the last case (*), $q_{max}$ is governed by the shadow transition.

| Case number | 1 | 2 | 3 | 4 | 5 |
|---|---|---|---|---|---|
| short wall description | smooth wall | Facetted wall | Pol. limiter (poloidally circular) | Pol. limiter (Poloidally flat segment) | Spherical modules |
| $q_{max}$ (MW/m²) | 0.6 | 2.6 | 3.2 | 3.2 | 4.8 |
| $q_{sep}^{//}$ (MW/m²) | 104.8 | 121.9 | 123 | 119 | 145 |
| $P_f$ | 7 | 38 | 37 | 37 | 40 |

Table 2 : Comparison of peak heat fluxes on various outer wall surfaces for $P_{total}$ = 15 MW.

| Mis-alignment (mm) | Power on bulging module (MW) | Peak heat flux (MW/m²) | Reg. heat flux (MW/m²) | Pf |
|---|---|---|---|---|
| 0 | 0.8 (15/18) | 0.71 | 0.71 | 3 |
| 2 | 1.0 | 1.5 | 0.71 | 6.3 |
| 5 | 1.6 | 2.8 | 0.68 | 11 |
| 10 | 3.0 | 5.0 | 0.61 | 20 |
| 20 | 6.0 | 9.1 | 0.45 | 38 |

Table 3 : Peak heat fluxes and peaking factors for increasingly bulging modules for $P_{total}$ = 15 MW.